# A Purely Classical Derivation of the Meissner Effect?

A.M. Gulian[*]

(Advanced Physics Laboratory, Chapman University, 15202 Dino Dr., Burtonsville, MD 20861, USA)

*Abstract.—* A recent study (*arXiv:1109.1968v2*, to be published in American Journal of Physics) claims to provide a classical explanation of the Meissner effect. However, the argument misuses de Gennes' derivation of flux expulsion in superconductors.

In a recent publication [1], Essén and Fiolhais attempt to explain the Meissner effect in superconductors in a "purely classical" way. They utilize many arguments, most of which we will ignore in this short remark. Rather, we address their most crucial argument based on the excerpts from de Gennes' classical textbook [2].

Following de Gennes, Ref. 1 writes the supercurrent density as

$$\boldsymbol{j}(\boldsymbol{r}) = n(\boldsymbol{r})e\boldsymbol{v}(\boldsymbol{r}), \qquad (1)$$

where $n$ is the density of superconducting electrons. In Ref. 1, they label $\boldsymbol{v}$ as the electron velocity, whereas De Gennes explicitly declares it to be carriers *drift*[1] velocity. Substituting this equation into the expression for kinetic energy

$$E_k = \int \frac{1}{2} n(\boldsymbol{r}) m \boldsymbol{v}^2(\boldsymbol{r}) dV, \qquad (2)$$

subject to $\nabla \times \boldsymbol{h} = (4\pi/c)\boldsymbol{j}$, and minimizing the sum of kinetic and magnetic energy

$$E_{mag} = \int (\boldsymbol{h}^2/8\pi) dV$$

with respect to the configuration of local magnetic field $\boldsymbol{h}$, one can arrive at the F. and H. Londons' equation:

$$\boldsymbol{h} + \lambda^2 \nabla \times (\nabla \times \boldsymbol{h}) = 0 \qquad (3)$$

where $\lambda$ is the penetration depth. Equation (3), of course, explains the field repulsion. Based on this, and also noticing that the derivation of Eq. (3) "utilizes no quantum concepts and contains no Planck constant," Essén and Fiolhais eventually deduce that superconductors "are just perfect conductors."

De Gennes himself never drew this conclusion. Moreover, he mentioned (in his remarks on the subject [2]), that in writing Eq. (1), one should "*assume the existence of permanent currents*" in superconductors. This statement helps to understand where the catch is in misusing Eq. (1). First of all, one should not consider the velocity $\boldsymbol{v}$ in Eq. (1) as just a redefinition of the current density $\boldsymbol{j}$ in the intuitive hydrodynamic sense, valid at the first glance, especially in view of the word "*drift*" for any system – otherwise the "Meissner effect" will immediately become a property of any metal, which is nonsense. Contrarily, it should be considered as a quantity characterizing the state of the system: its kinetic energy. For superconductors, we have a *permanent current* (*v* is constant in time) in response to the presence of magnetic field. Correspondingly, a decaying current *v* = *v(t)* → *0* should be supposed for a viscous flow.

---

[*] gulian@chapman.edu
[1] Hereafter the words in cursive belong to de Gennes [2].

Having noticed that, we can now distinguish two cases: 1) the viscosity is removed first, and then the magnetic field is turned on; 2) the magnetic field is on when the metal is still in the viscous state, and only after the relaxation of the induced current the viscosity is removed. In case 1, the action of the field on the ideal conductor is dynamical, and it sets up Eddy currents which, in equilibrium, are distributed in accordance to Eq. (3). This picture is based on classical concepts and comes close to the experimental findings. However, in case 2, the velocity *v(r, t)* first comes to zero for all the values of *r*, and stays zero forever after the viscosity is removed[2]. As follows from equations (1)-(3) with *v=0*, there is no redistribution of the applied magnetic field. This contradicts to the experimental observations for superconductors, first made by Meissner and Ochsenfeld [4].

Curious readers may exclaim at this point: for a given magnetic field, how does the superconductor "know" that it should hold a finite value of *v*, and the normal metal does not? The answer is that the quantum objects, such as superconductors, can "feel" the presence of the vector potential *A*, so that an alternative form of the Londons' equation is

$$\boldsymbol{j} = -\frac{e^2 n}{m} \boldsymbol{A}, \qquad (4)$$

(here m is the mass of a charge carrier). "Ideal" metals rather obey the relation

$$\boldsymbol{j} = \sigma \boldsymbol{E}, \qquad (5)$$

which, to be finite in the limiting case of conductivity $\sigma \to \infty$, requires electric field $\boldsymbol{E} \to 0$. Even then, we have uncertainty of $\infty \cdot 0$ –type. An attempt to assign a limiting value to Eq. (5) coinciding with Eq. (4) will immediately bring in a contradiction in case of ideal metals: in classical physics, observables cannot be proportional to the vector-potential. Situation is different in superconductors, where the (Bose-condensed) charge carriers have a macroscopic wave function

$$\Psi = \sqrt{n} \exp(i\Theta) \qquad (6)$$

which restores the gauge-invariance of the equation (4):[3]

$$\boldsymbol{j} = -\frac{\hbar e n}{m} \left( \nabla \Theta - \frac{e}{\hbar c} \boldsymbol{A} \right). \qquad (7)$$

---

[2] The fact that "ideal" conductors should hold in the state in which the removal of viscosity took place follows from classical electrodynamics and is discussed in detail in F. London's book [3].
[3] At this stage the Planck's constant enters the theory, though the quantum description started earlier with Eq. (4), and, actually, even with Eq. (1) by suggesting *v* ≠ *0*.

The ideal metal falls short with this issue. Despite each of the carriers may have its own wave function, their phases are not correlated on the macroscopic scale. Therefore, the ideal metals are not able to "*find an equilibrium state where the sum of kinetic and magnetic energies is minimum*" whereas for superconductors, this takes place, *"and this state, for macroscopic samples, corresponds to the expulsion of the magnetic field "*.

The majority of physicists have grown up with classical textbooks, such as [5], and traditionally believe that the Meissner effect inherently reveals the quantum beauty of superconductivity. Although the majority does not determine the truth in science, there is no need to change the traditional point of view at this time: by its entirety, Meissner effect cannot be explained classically.

Acknowledgement: I would like to express my deep gratitude to P. Abramian Barco, D. Van Vechten, S. Nussinov and M. Gulian for critical reading of this manuscript.